\documentclass{fmj2010}
\usepackage{graphicx}
\usepackage{color}
\def\fermilat{\textit{Fermi}/LAT}
\def\fermi{\textit{Fermi}}

\begin{document}
\title{Multi-band properties of superluminal AGN detected by \fermilat\ }
%\title{The optical continumm emission in superluminal AGN detected by

   \author{ T.G. Arshakian\inst{1}\thanks{Speaker}
           \and
           J. Le\'on-Tavares \inst{2}
           \and
          J. Torrealba \inst{3}
          \and
          V.H. Chavushyan \inst{4}
%          \and
%          et al. 
          }

   \institute{ Max-Planck-Institut f\"ur Radioastronomie, Auf dem H\"ugel 69,
   53121 Bonn, Germany\\
   \email{t.arshakian@mpifr-bonn.mpg.de}
          \and
 Aalto University Mets\"ahovi Radio Observatory, Mets\"ahovintie 114, FIN-02540, Kylm\"al\"a, Finland\\
   \email{leon@kurp.hut.fi}
         \and
   Instituto de Astronom\'{\i}a, Universidad Nacional Aut\'onoma de M\'exico, Apartado Postal 70-264,
   04510 M\'exico D.F., M\'exico, \email{cjanet@astroscu.unam.mx}
       \and
   Instituto Nacional de Astrof\'{\i}sica \'Optica y
   Electr\'onica, Apartado Postal 51 y 216, 72000 Puebla, Pue, M\'exico\\
   \email{vahram@inaoep.mx}
  }

  \abstract{We perform a multi-band statistical analysis of core-dominated superluminal active galactic nuclei (AGN) detected with \fermi\ Large Area Telescope (LAT). The detection rate of $\gamma$-ray jets is found to be high for optically bright AGN. There is a significant correlation between the $\gamma$-ray luminosity and the optical nuclear and radio (15~GHz) luminosities of AGN. We report a well defined positive correlation between the $\gamma$-ray luminosity and the radio-loudness for quasars and BL Lacertae type objects (BL~Lacs). The slope of the best-fit line is significantly different for quasars and BL~Lacs. The relations between the optical and radio luminosities and the $\gamma$-ray loudness are also examined, showing a different behavior for the populations of quasars and BL~Lacs. Statistical results suggest that the $\gamma$-ray, optical and radio emission is generated at different locations and velocity regimes along the parsec-scale jet.}

   \maketitle
%
%________________________________________________________________

\section{Introduction}

Extreme optical variability was a clear signature for some blazars
detected at $\gamma$-ray energies during the EGRET era (Fitchel et
al. 1994). Using the \fermilat\ improved sensitivity, recent
multiwavelength variability studies on individual sources have
confirmed the tight connection among the $\gamma$-ray and optical
variable emission (e.g. 3C~273 in Abdo et al. \cite{abdo10a}). So far there was 
no attempt to compare optical continuum properties among
$\gamma$-ray blazars. In this manuscript we take advantage of the
unprecedented sensitivity provided by the first bright source
catalogue (1FGL; Abdo et al. \cite{abdo10b}) to study the multi-band correlations between optical, radio, and $\gamma$-ray emission in superluminal AGN detected by \fermilat\ during its first
11 months of operation. We adopt a flat 
cosmology with $H_{0} = 71$ km s$^{-1}$ Mpc$^{-1}$, $\Omega_{m} = 0.27$, and $\Omega_{\Lambda} = 0.73$.

 \begin{figure}
\includegraphics[width=\columnwidth]{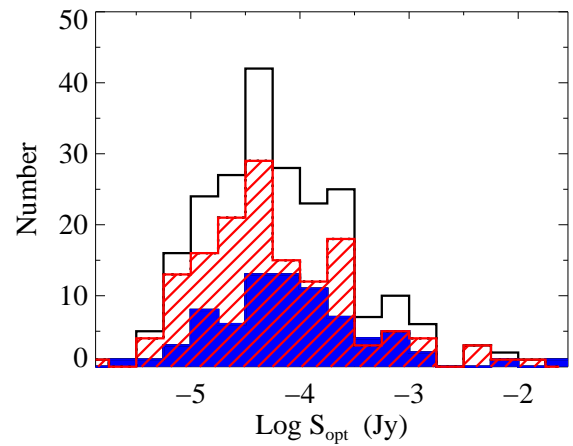}
\caption{Distributions of optical nuclear fluxes for AGN from the MOJAVE sample (full line), 102 AGN detected as the gamma-ray source (filled histogram) and non-detected with \fermilat\ during
  the 1FGL period (shaded histogram).}
\end{figure}

\section{The sample of AGN detected with \fermilat}
The MOJAVE-2 sample consists of about 290 core-dominated AGN from the MOJAVE (Monitoring of Jets 
in AGN with VLBA Experiments) program (Lister et al. 2009). The sample of AGN studied in this manuscript includes 250 core-dominated AGN (Kovalev et al. 2005) from the MOJAVE-2 sample. Most of these AGN are currently monitored with the Very Large Baseline Array (VLBA) at 15~GHz. The 1FGL catalogue
includes about 190 sources from the MOJAVE-2 sample. Optical nuclear fluxes and redshift measurements were available for a sample of 102 MOJAVE sources (Arshakian et al. \cite{arshakian10a}) identified by
\fermilat\ (hereafter, M-1FGL sample). Seventy six out of 102 sources in the M-1FGL sample
are part of the statistically complete MOJAVE-1 sample. We will refer to these 76 sources as the M1-1FGL sample. The Kolmogorov-Smirnov (K-S) test shows  
that the distribution of redshifts in the MOJAVE-1 sample and the
M1-1FGL sample are drawn from the same parent population indicating that the M1-1FGL sample is not biased by redshift.

The M-1FGL sample consists of 76 quasars, 24 BL~Lacs, and two radio
galaxies which are excluded from further statistical tests. We use the non-parametric Kendall's $\tau$ test to analyze correlations between independent variables and the partial Kendall's $\tau$ test to account for the common dependence on redshift in the correlations between luminosities. Throughout the paper we assume a correlation to be significant if a chance probability $P < 0.05$.

\begin{figure}
 \includegraphics[width=\columnwidth]{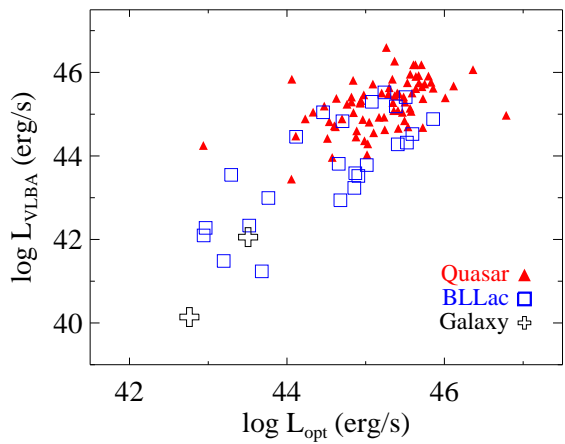}
 \caption{Radio luminosity at 15~GHz against optical
   nuclear luminosity at 5100\,\AA.}
\end{figure}

\section{Interplay between $\gamma$-ray, optical, and radio properties of superluminal AGN}

The distribution of optical nuclear fluxes for the MOJAVE AGN detected and non-detected with \fermilat\ is presented in Figure~1.
At first glance, the \fermilat\ detected sources seem to have higher optical fluxes than those with no-detection.
This is further supported by the K-S statistical test: there is a  
significant difference (at a confidence level of $99.9$\,\%) between the distributions of optical fluxes of AGN detected and non-detected by \fermilat. When using the M1-1FGL sample, the difference is significant at a confidence level of 96\,\%. We conclude that the detection rate of $\gamma$-ray AGN is high for sources having high optical nuclear fluxes.

\begin{figure}
 \includegraphics[width=\columnwidth]{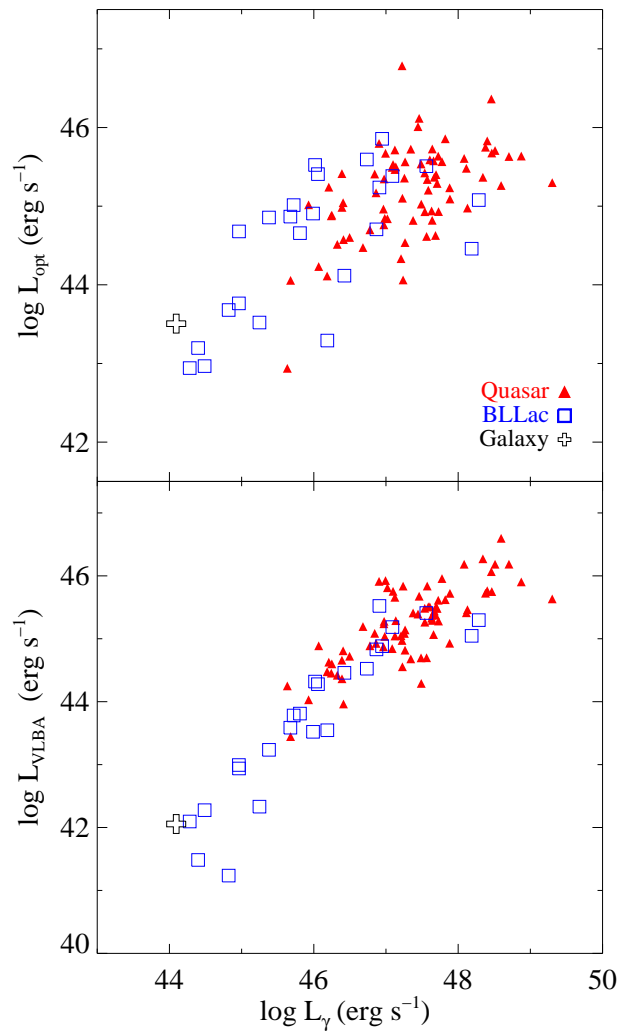}
 \caption{Optical nuclear luminosity against $\gamma$-ray luminosity (top panel), and radio (15~GHz) luminosity against $\gamma$-ray luminosity  
   (bottom panel) for AGN from the M-1FGL sample. Labels in the top panel denote the corresponding population. }
\end{figure}
We derive the optical nuclear luminosities ($L_{\rm opt}$), total VLBA luminosities ($L_{\rm VLBA}$), and the rest-frame radio-loudness ($R\propto S_{\rm VLBA}/S_{\rm opt}$) using the fluxes
given in Arshakian et al. (\cite{arshakian10a}). $\gamma$-ray
luminosities were computed using the Eq.~(1) in Ghisellini et
al. (2009), where $S_{\gamma} (\nu_{1}, \nu_{2})$ is the energy flux between 0.1~GeV and 100~GeV from the 1FGL catalogue. Note that the $\gamma$-ray, optical, and radio luminosities are estimated from non-simultaneous observations.

\emph{Optical--Radio emission.} Arshakian et al. (\cite{arshakian10a}) found a positive correlation in the $L_{\rm opt}-L_{\rm VLBA}$ relation plane for AGN from the MOJAVE-1 sample. They concluded that the correlation is due to the population of quasars and that the optical emission is non-thermal and generated in the parsec-scale jet.
This is supported by studies of individual radio galaxies, 3C\,390.3 and 3C\,120 (Arshakian et al. \cite{arshakian10b} and Le\'on-Tavares et al. \cite{tavares10}) for which the link between optical continuum variability and kinematics
of the parsec-scale jet was found. It was interpreted in terms of optical continuum 
flares generated at subparsec-scales in the innermost part of a relativistic jet rather than in the accretion disk. 
We confirm the $L_{\rm opt}-L_{\rm VLBA}$ positive correlation for a larger sample of M-1FGL quasars as well as no-correlation for BL~Lacs (see Figure 2 and Tables~1 and 2).
\begin{figure}
 \includegraphics[width=\columnwidth]{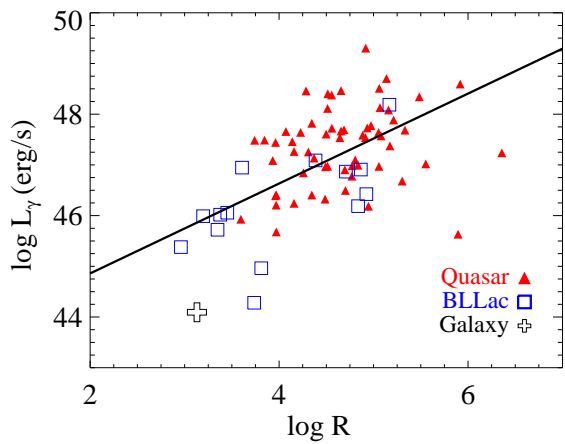}
 \caption{$\gamma$-ray luminosity ($L_{\gamma}$) versus radio-loudness ($R$) for M-1FGL AGN. The solid line represents the ordinary least-square fit to the data.}
\end{figure}

\emph{Gamma-ray--Optical emission.} We find a positive correlation between $L_{\gamma}$ and $L_{\rm opt}$ (Figure~3). The correlation is significant for quasars from the M-1FGL and M1-1FGL samples and it is stronger for M1-1FGL quasars (Tables~1 and 2), suggesting for a single production mechanism for $\gamma$-ray and optical nuclear emission.

\emph{Gamma-ray--Radio emission.} Kovalev et al. (2009) reported a significant correlation between $\gamma$-ray and radio VLBA (8~GHz) emission for sample of $\sim$30 AGN. This correlation holds at high confidence level ($>99.9$\,\%; Figure~3) for non-simultaneous measurements and a larger sample (M-1FGL) of AGN. This suggests that the powers averaged over the long time scales are correlated and, hence, the Doppler-factors of the parsec-scale jet in the gamma and radio domains are not changing substantially on a timescale of a few years. 
Pushkarev et al. (this proceedings) found that the $\gamma$-ray emission leads the radio emission of the parsec-scale jet at 15~GHz with time delay of few months. They interpreted the observed time lag as a result of synchrotron opacity in the jet: the radio and $\gamma$-ray emission are generated in the same region (perturbation in the jet?) and become observable with some time delay due to the opacity effects. If this scenario is correct then the variable optical emission is also generated in the perturbation moving upstream the jet, and  we should expect that the optical emission leads the radio emission and delays with respect to the $\gamma$-ray emission. 
%\\

\begin{figure*}
 \includegraphics[width=\textwidth]{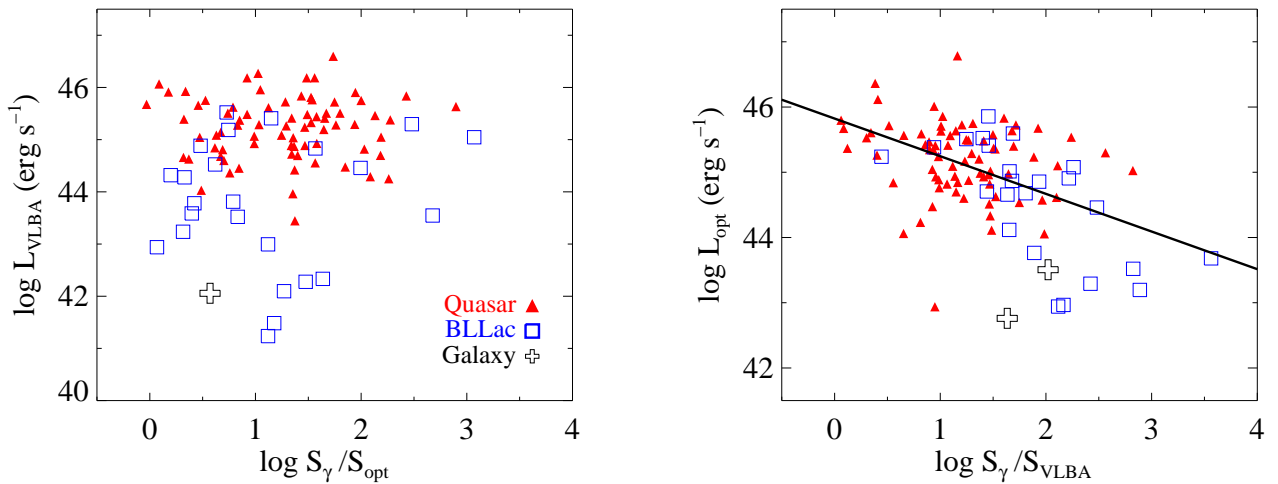}
 \caption{Radio VLBA luminosity against $\gamma$-optical loudness (left panel), and
optical luminosity against  $\gamma$-radio loudness  (right-panel) for the M-1FGL quasars and BL~Lacs. The solid line represents the ordinary
   least-square fit to the data.}
\end{figure*}

The correlation in the $L_{\rm VLBA}-L_{\gamma}$ relation plane is apparently stronger than that in the $L_{\rm opt}-L_{\gamma}$ (Figure~3). 
%For a given $L_{\gamma}$, the range of optical nuclear luminosities is larger than the range of VLBA luminosities. 
One may think that the corrected optical nuclear emission of some AGN (Arshakian et al. \cite{arshakian10a}) might be contaminated by contribution of a stellar component thus causing the large dispersion in the $L_{\rm opt}-L_{\gamma}$ diagram. Contamination should be stronger in  
radio galaxies and weaker in quasars and BL~Lacs for which the contribution of optical nuclear emission is dominant. The large dispersion in the $L_{\rm opt}-L_{\gamma}$ relation plane can be due to non-simultaneous optical/$\gamma$-ray observations and stronger variability in the optical regime than that in the radio, and/or wider range of Doppler factors in the optical regime compared to the range of Doppler factors of the jet at 15~GHz, if the bulk of optical emission is generated in the relativistic jet and it is Doppler boosted.

We report a significant positive correlation ($> 99\,\%$) between $L_{\gamma}$ and
radio-loudness for quasars and BL Lacs (see Figure~4 and Table~1). The solid line in Figure~5 represents the best fit to the data, 
\begin{equation}
 \log L_{\gamma} =   (0.95\pm 0.11)\log R +  (42.76 \pm 0.25). 
\end{equation}
The $L_{\gamma} \propto R$ relation suggests that the strong $\gamma$-ray jets have progressively high Doppler factors (or faster speeds) in the radio domain compared to those in the optical regime. For quasars, the regression line is fitted by 
\begin{equation}
 \log L_{\gamma}=(0.33 \pm 0.14)  \log R +(45.8 \pm 0.32),
\end{equation}
while for BL~Lacs the best fit is,
\begin{equation}
 \log L_{\gamma} =   (0.98\pm 0.21)\log R +  (42.29 \pm 0.44).
\end{equation}
It is noticeable that the slope derived for quasars is shallower than the slope fitted for BL~Lacs. 
The $L_{\gamma}-R$ correlation is still significant for all AGN ($> 99$\,\%), quasars and BL~Lacs ($99\,\%$) of the M1-1FGL sample (Table~2). The best-fit parameters for the later sample are almost unchanged. 

\begin{table}[ht]
  \caption[]{Kendall's $\tau$ correlation analysis between emission characteristics of AGN from the M-1FGL sample. A1 and A2 are the independent variables for which the KendallÕs $\tau$ correlation 
analysis is performed,  $\tau$ is the correlation coefficient, and $P$ is the 
probability of a chance correlation. The correlations are considered to be significant if the chance probability $P < 0.05$ (or confidence level $>95$\,\%).}

{\tiny

\begin{tabular}{cccccccccccccc}
 \hline \hline
   &  & \multicolumn{2}{c}{All} &  &  \multicolumn{2}{c}{Quasars} & & \multicolumn{2}{c}{BL Lacs}\\
 \cline{3-4}\cline{6-7}\cline{9-10} \smallskip
  A1 & A2   &   $\tau$ & $P$ &  &  $\tau$ & $P$ & & $\tau$ & $P$ \\
 \hline\hline
$ L_{\rm VLBI}$ & $ L_{\rm opt}$  &  0.2 & 7e-3& &0.2 &2e-2& &0.1 & 0.4\\
$ L_{\rm \gamma}$ & $ L_{\rm opt}$  &  0.1 & 4e-2& & 0.1 & 5e-2& &0.1 & 0.3\\
$ L_{\rm \gamma}$ & $ L_{\rm VLBA}$  &  0.4 & 6e-8& &0.3 & 7e-5& &0.5 & 7e-4\\
$ L_{\rm \gamma}$ &  $R$                     &  0.4 & 3e-9& &0.2 & 8e-3& &0.4 & 1e-4\\
$ L_{\rm opt}$   & $S_{\gamma}/S_{\rm VLBA}$ &  -0.3 & 1e-5& &0.2 & 4e-2& &0.5 & 1e-3\\

\end{tabular}
}
\end{table}

\begin{table}[h]
  \caption[]{Kendall's $\tau$ correlation analysis between emission characteristics of AGN from the M1-1FGL sample. }

{\tiny

\begin{tabular}{cccccccccccccc}
 \hline \hline

   &  & \multicolumn{2}{c}{All} &  &  \multicolumn{2}{c}{Quasars} & & \multicolumn{2}{c}{BL Lacs}\\
 \cline{3-4}\cline{6-7}\cline{9-10} \smallskip
  A1 & A2   &   $\tau$ & $P$ &  &  $\tau$ & $P$ & & $\tau$ & $P$ \\
 \hline\hline
$ L_{\rm VLBA}$ & $ L_{\rm opt}$    &  0.2 & 1e-2& &0.2 & 2e-2& &0.1 & 0.6\\
$ L_{\rm \gamma}$ & $ L_{\rm opt}$  &  0.2 & 2e-2& &0.2 & 1e-2& &0.1 & 0.7\\
$ L_{\rm \gamma}$ & $ L_{\rm VLBA}$ &  0.4 & 1e-6& &0.3 & 1e-4& &0.4 & 5e-2\\
$ L_{\rm \gamma}$ & $R$      &  0.4 & 3e-6& &0.2 & 1e-2& &0.6 & 1e-2\\
$ L_{\rm opt}$   & $S_{\gamma}/S_{\rm VLBA}$ &  -0.3 & 1e-4& &0.2 & 4e-2& &0.5 & 2e-3\\

\end{tabular}
}
\end{table}

\emph{Gamma-ray loudness.} We define the $\gamma$-optical loudness and $\gamma$-radio loudness as the $S_{\gamma}/S_{\rm opt}$ and $S_{\gamma}/S_{\rm VLBA}$, respectively. We find that the radio luminosity is independent of $\gamma$-optical loudness (Figure~5, left panel). On the other hand, there is a negative correlation in the $L_{\rm opt}-S_{\gamma}/S_{\rm VLBA}$ relation plane (Figure~5, right panel), which is much stronger for BL Lacs (confidence level $\ga 99.9\,\%$; Tables~1 and 2). 
For the M-1FGL AGN, the best-fit regression line is represented by $L_{\rm opt} \propto (S_{\gamma}/S_{\rm VLBA})^{-0.57}$. We interpret this relation as the optically weaker jets to have higher Doppler factors (or Lorentz factors if outflows radiating in optical and $\gamma$-ray bands have the same viewing angle) in the gamma domain compared to those in the radio domain.  
The slope of the best-fit line is steeper for BL~Lacs ($-0.89\pm0.21$) than for quasars ($-0.24\pm0.1$) and the difference is statistically significant. We suggest that, on the average, the ratio between the Doppler factors in gamma and radio regimes varies slower for quasars than for BL~Lacs, indicating for different velocity regimes in their jets. 
%\\

The significant differences found between quasars and
BL~Lacs are supported in recent studies (e.g. Ghisellini et al. 2009,
Sambruna et al. 2010, Tornikoski et al., in this proceedings) suggesting the presence of different
physical conditions along the jet in quasars and BL~Lacs. 
This and other insights into the relationship between the $\gamma$-ray, radio, and optical emission will be pursued in a more detail in further studies.

\section{Summary}
Using the sample of 100 superluminal quasars and BL~Lacs detected by \fermilat, we
investigate relations between their optical, radio, and $\gamma$-ray
emission available from non-simultaneous observations. 
Our main results are summarized as follows:

\begin{itemize}
\item The detection rate of $\gamma$-ray superluminal AGN is high for optically bright AGN. 

\item The known positive correlation between $L_{\rm opt}$ and $L_{\rm VLBA}$ holds for the M-1FGL quasars at a confidence level of $98\,\%$. The known correlation between $\gamma$-ray flux and radio flux density (measured quasi-simultaneously) is also valid for non-simultaneous  measurements of $L_{\gamma}$ and $L_{\rm VLBA}$ for quasars. The $L_{\gamma}-L_{\rm VLBA}$ correlation is significantly stronger than that in the $L_{\gamma}-L_{\rm opt}$ relation plane.
%It is most likely that the optical emission is non-thermal and generated in the innermost part of the parsec-scale jet. 

\item There is a correlation between $L_{\rm opt}$ and $L_{\gamma}$ which exclusively holds for quasars. The correlation is significant for the M-1FGL quasars (at a confidence level of $99\,\%$) and marginally significant for quasars from the M1-1FGL sample (c.l. $95\,\%$).

\item We report a statistically significant positive correlation (c.l. $>99\,\%$) between $\gamma$-ray luminosity and radio-loudness for both, quasars and BL Lacs. The slope of the $L_{\gamma}-R$ relation is found to be steeper for the population of BL Lacs.

\item We find that the radio luminosity at 15~GHz is independent of the $\gamma$-radio loudness ($S_{\gamma}/S_{\rm VLBA}$) for quasars and BL Lacs.
The $\gamma$-optical loudness ($S_{\gamma}/S_{\rm opt}$) and optical nuclear luminosity are negatively correlated for quasars (c.l. $96\,\%$) and BL Lacs (c.l. $99.9\,\%$) with the slope of the later being steeper. 

\end{itemize}

\begin{acknowledgements}
  TGA acknowledges support by DFG-SPP project under grant 566960. This work was supported by CONACYT research grant 54480
  (Mexico). 
\end{acknowledgements}

\end{document}